\documentclass[10pt]{article}

\usepackage{geometry}
\usepackage{amsmath}
\usepackage{amssymb}
\usepackage{amsthm}
\usepackage{mathrsfs}
\usepackage{enumerate}
\usepackage{epsf}
\usepackage{psfrag}
\usepackage{dsfont}
\usepackage{subcaption}
\usepackage{graphicx}
\usepackage{empheq}
\usepackage{xcolor}
\usepackage{authblk}
\usepackage{url}
\usepackage[linktocpage]{hyperref}
\usepackage{cleveref}

\hypersetup{
	colorlinks,
	linkcolor={red!60!black},
	citecolor={green!50!black},
	urlcolor={blue!80!black}
}

\DeclareGraphicsExtensions{.eps,.art,.ART,.ps}
\allowdisplaybreaks
\newcommand{\cL}{\mathcal{L}}

\newcommand{\bbR}{\mathbb{R}}      

\makeatletter
\newcommand*{\defeq}{\mathrel{\rlap{%
\raisebox{0.3ex}{$\m@th\cdot$}}%
\raisebox{-0.3ex}{$\m@th\cdot$}}%
=}
\makeatother
\begin{document}
\title{Relativistic elastic membranes: rotating disks and Dyson spheres}
\author[1]{Paulo Mour\~{a}o}
\author[1]{Jos\'e Nat\'ario}
\author[2]{Rodrigo Vicente}
\affil[1]{\normalsize CAMGSD, Departamento de Matem\'{a}tica, Instituto Superior T\'{e}cnico, Universidade de Lisboa, Portugal}
\affil[2]{\normalsize Institut de Fisica d'Altes Energies (IFAE), The Barcelona Institute of Science and Technology, \qquad Campus UAB, 08193 Bellaterra (Barcelona), Spain}
\date{}
\maketitle
\begin{abstract}
We derive the equations of motion for relativistic elastic membranes, that is, two-dimensional elastic bodies whose internal energy depends only on their stretching, starting from a variational principle. We show how to obtain conserved quantities for the membrane's motion in the presence of spacetime symmetries, determine the membrane's longitudinal and transverse speeds of sound in isotropic states, and compute the coefficients of linear elasticity with respect to the relaxed configuration. We then use this formalism to discuss two physically interesting systems: a rigidly rotating elastic disk, widely discussed in the context of Ehrenfest's paradox, and a {\em Dyson sphere}, that is, a spherical membrane in equilibrium in Schwarzschild's spacetime, with the isotropic tangential pressure balancing the gravitational attraction. Surprisingly, although spherically symmetric perturbations of this system are linearly stable, the axi-symmetric dipolar mode is already unstable. This may be taken as a cautionary tale against misconstruing radial stability as true stability.
\end{abstract}
\tableofcontents
%
%
%
%
%
\section{Introduction}
The requirement that no causal influence can propagate faster than the speed of light greatly constrains fully relativistic models of extended bodies. Among these, the simplest are arguably those obtained by employing the theory of relativistic elasticity, which was formulated in its modern form by Carter and Quintana in \cite{CQ72} (see also \cite{BS03, Beig23} for more recent accounts). This theory has been used extensively to construct simplified versions of astrophysical objects \cite{KS03, KS04, ABS09, BS08, AOS16, BS17, ANPR22a, ANPR22b, ANPR24a, ANPR24b}, and is expected to play an important role in modeling neutron star crusts \cite{CH08, AHCS19}.

Unfortunately, the highly nonlinear equations of relativistic elasticity are quite difficult to handle, which limits their usefulness in providing simple motions of extended objects. Some simplification is achieved by considering one-dimensional elastic objects; this approach was taken in \cite{NQV18}, where the motion of rotating elastic string loops around black holes was studied in connection with stability, cosmic censorship and the Penrose process (see also \cite{Natario14, MNS24, NSV24}). In the current paper we take the next logical step and discuss relativistic elastic membranes, that is, two-dimensional elastic bodies whose internal energy depends only on their stretching, first studied by Carter in a series of papers \cite{Carter92, Carter94, Carter01, Carter03, Carter07, Carter11} (see also \cite{KF08, AO13}). We find that the theory becomes considerably more intricate than that of elastic strings, reflecting the fact that the geometry of a surface is much richer than that of a curve. This includes not only aspects similar to (albeit more complex than) those discussed for elastic strings in \cite{NQV18}, such as the variational derivation of the equations of motion, conserved quantities and the speeds of sound, but also entirely new features with no analogue for elastic strings, such as the Poisson ratio or the bulk modulus. After systematically developing the theory of relativistic elastic membranes, we illustrate its application in a couple of physically interesting systems: a rigidly rotating disk in the Minkowski spacetime, and a static spherical membrane in the Schwarzschild spacetime.

The organization of the paper is as follows: in Section \ref{section1} we derive the equations of motion for relativistic elastic membranes starting from a Lagrangian density, and rewrite these equations as conservation of energy-momentum along the membrane's {\em worldtube} (that is, the $3$-dimensional submanifold of spacetime traced out by the membrane) plus the so-called {\em generalized sail equations} (the vanishing of the contraction between the energy-momentum tensor and the extrinsic curvature of the worldtube). We also show how to obtain conserved quantities for the membrane's motion from spacetime symmetries, determine the membrane's longitudinal and transverse speeds of sound in isotropic states, and compute the coefficients of linear elasticity with respect to the relaxed configuration. In Section \ref{section2} we give an example of a simple $1$-parameter family of elastic laws with longitudinal speed of sound equal to the speed of light, which we dub the {\em rigid membrane}, and use it to construct, for the first time, explicit examples of rigidly rotating elastic disks, widely discussed in the context of Ehrenfest's paradox \cite{Ehrenfest09, Gron04}. In Section \ref{section3} we consider a {\em Dyson sphere}, that is, a spherical membrane in equilibrium in Schwarzschild's spacetime. We obtain the equilibrium configuration, where the gravitational attraction is balanced by the isotropic tangential pressure, and analyze its linear stability. Surprisingly, although the spherically symmetric mode is stable (yielding what is usually called a {\em breathing mode}), the axi-symmetric dipolar mode is already unstable. This may be taken as a cautionary tale against misconstruing radial stability as true stability.

We follow the conventions of \cite{MTW73, W84}; in particular, we use a system of units for which $c=G=1$. Greek letters $\alpha, \beta, \ldots$ represent spacetime indices, running from $0$ to $n$, small case Latin letters $i, j, \ldots$ represent spatial indices, running from $1$ to $n$ (or sometimes from $3$ to $n$), and capital Latin letters $A, B, \ldots$ represent indices in the membrane's worldtube, running from $0$ to $2$. We used {\sc Mathematica} for symbolic and numerical computations, and also to produce various plots.
%
%
%
\section{Elastic membrane theory}\label{section1}
In this section, we use a variational approach to (re-)derive the equations of motion of a relativistic elastic membrane, for the reader's convenience and also to fix notation (see \cite{Carter92, Carter01, Carter11} for alternative derivations). These equations are then shown to be equivalent to the conservation of an energy-momentum tensor defined on the worldtube plus the vanishing of the contraction between the energy-momentum tensor and the extrinsic curvature of the worldtube (dubbed the {\em generalized sail equation} in \cite{CM93}). We obtain the conserved quantities associated to Killing vector fields (as required by Noether's theorem), and compute the coefficients of linear elasticity with respect to the relaxed configuration. Finally, we determine the membrane's longitudinal and transverse speeds of sound in an isotropic state.
\subsection{Lagrangian density}
We define a {\em membrane} as a two-dimensional elastic body whose internal energy depends only on its stretching; it is, of course, an idealization, and should be regarded as an approximation of a three-dimensional elastic body whose thickness is much smaller than its width, and whose bending energy can be disregarded. We model a membrane moving on a $(n+1)$-dimensional spacetime $(M,g)$ by an embedding $X\colon \mathbb{R}\times S\to M$, where $S$ is a $2$-dimensional manifold labeling the points of the membrane. We assume that $S$ carries a Riemannian metric, determining the membrane's relaxed configuration, which can always be written in local Gaussian coordinates $(\lambda^1, \lambda^2)$ as
\begin{equation}
k = \left(d\lambda^1\right)^2 + \left(f\left(\lambda^1,\lambda^2\right)\right)^2\left(d\lambda^2\right)^2.
\end{equation}
The curve $\tau \mapsto X(\tau, \lambda^1, \lambda^2)$ is then the worldline of the point of the membrane labeled by $(\lambda^1,\lambda^2)$.

The embedding $X$ induces a Lorentzian metric
\begin{equation}\label{metric_embedd}
h_{AB}=g_{\mu\nu}(X)\partial_AX^\mu\partial_BX^\nu
\end{equation}
on $\mathbb{R}\times S$, which we identify with its image $\Sigma=X(\mathbb{R}\times S)$ (sometimes called the membrane's \textit{worldtube}).
Choosing a local orthonormal frame $\{E_0,E_1,E_2\}$ tangent to $\Sigma$ such that $E_0$ is the $4$-velocity of the membrane's particles, we must have
\begin{equation}
\left\{
\begin{array}{lll}
\dfrac{\partial X}{\partial \tau}=\alpha E_0\\[8pt]
\dfrac{\partial X}{\partial \lambda^1}=\beta_1 E_0+\sigma_{11}E_1+\sigma_{21}E_2\\[8pt]
\dfrac{\partial X}{\partial \lambda^2}=\beta_2 E_0+f\sigma_{12}E_1+f\sigma_{22}E_2
\end{array}
\right. \,,
\end{equation}
for some smooth locally defined functions $\alpha, \beta_1, \beta_2, \sigma_{11},\sigma_{12},\sigma_{21},\sigma_{22}$. Note that the matrix
\begin{equation}
\boldsymbol{\sigma}\equiv \begin{pmatrix}
\sigma_{11} & \sigma_{12}\\
\sigma_{21} & \sigma_{22}
\end{pmatrix}
\end{equation}
represents the linear deformation of the membrane\footnote{More precisely, the derivative of the embedding restricted to a hypersurface of constant $\tau$, composed with the orthogonal projection on the simultaneity hyperplanes.} according to an observer comoving with it, written in the orthonormal basis $\{\partial_{\lambda^1}, (1/f) \partial_{\lambda^2}\}$ of the relaxed configuration and the orthonormal basis $\{E_1, E_2\}$ of the hyperplane of simultaneity of such an observer. In particular, the infinitesimal distances between points in the deformed membrane are described by the matrix $\boldsymbol{\sigma}^T \cdot \boldsymbol{\sigma}$.

The components of the metric induced on the worldtube are
\begin{equation}
(h_{AB})=-\begin{pmatrix}
\alpha^2 & \alpha\,\boldsymbol{\beta}^T \\
\alpha\,\boldsymbol{\beta} & \boldsymbol{\beta}\cdot \boldsymbol{\beta}^T-\boldsymbol{\phi}^T \cdot \boldsymbol{\sigma}^T \cdot \boldsymbol{\sigma} \cdot \boldsymbol{\phi} & 
\end{pmatrix}\,,
\end{equation}
where
\begin{equation*}
\boldsymbol{\phi}\equiv \begin{pmatrix}
1 & 0 \\
0 & f
\end{pmatrix} \,,
\end{equation*}
and so
\begin{equation}
h \equiv \det(h_{AB})=-\alpha^2 f^2 \det\left(\boldsymbol{\sigma}^T\cdot\boldsymbol{\sigma}\right)=h_{00} \det\left(\boldsymbol{\sigma}^T\cdot\boldsymbol{\sigma}\right)\,.
\end{equation}
Defining the \textit{number density} $n\equiv 1/\sqrt{\det\left(\boldsymbol{\sigma}^T\cdot \boldsymbol{\sigma}\right)}$ and the \textit{shear} $s\equiv \sqrt{{\rm tr}\left(\boldsymbol{\sigma}^T\cdot \boldsymbol{\sigma}\right)}$, we have
\begin{equation}\label{n2s2}
\left\{
\begin{array}{ll}
\displaystyle n^2=\frac{h_{00}f^2}{h}\\[8pt]
\displaystyle s^2=h_{11}+\frac{h_{22}}{f^2}-\frac{1}{h_{00}}\left(h_{01}^2+\frac{h_{02}^2}{f^2}\right)
\end{array}
\right.\,.
\end{equation}
To obtain the membrane's equations of motion we must choose an action 
\begin{equation}
S=\int_{\mathbb{R}\times I} \mathcal{L}(X,\partial X)\, d\tau d\lambda^1 d\lambda^2\,.
\end{equation}
For an isotropic elastic membrane, whose internal energy density $\rho$ must be of the form $\rho=F(n^2,s^2)$, the Lagrangian density is 
\begin{equation*}
\mathcal{L}=F(n^2,s^2)\sqrt{-h}\,,
\end{equation*}
where $h$, $n^2$ and $s^2$ are given as functions of $\left(X,\partial X\right)$ by equations~\eqref{metric_embedd} and~\eqref{n2s2}. In the particular case when $F$ is constant we obtain the action for Nambu-Goto membranes (or $2$-branes); this action is proportional to the worldtube's volume, and consequently has special properties, such as reparameterizarion invariance.

\subsection{Equations of motion}

To find the equations of motion we must compute the variation $\delta \mathcal{L}$ of the Lagrangian density resulting from a variation $\delta X$ of the embedding. Using the well-known formula for the variation of the determinant of the metric,
\begin{equation}
\delta h=h h^{A B}\delta h_{A B} \,,
\end{equation}
we obtain\footnote{We represent by $F^{(i,j)}$ the mixed partial derivative of the function $F$ of order $i$ with respect to the first variable and order $j$ with respect to the second variable.}
\begin{align} \label{deltaL}
\delta \mathcal{L}= \left(\frac{1}{2}F \sqrt{-h}\, h^{A B}+F^{(1,0)}\frac{h_{00}h^{A B}-\delta_0^A \delta_0 ^B}{\sqrt{-h}}f^2+F^{(0,1)}\sqrt{-h}\,k^{A B}\right)\delta h_{A B}\,,
\end{align}
where
\begin{align}
k^{AB} & \equiv
\delta_1^A \delta_1^B+\frac1{f^2}\delta_ 2^A \delta_2^B+\frac{\delta_0^A \delta_0^B}{h_{00}^2}\left(h_{0 1}^2+\frac{h_{0 2}^2}{f^2}\right)-\frac{1}{h_{00}}\left(h_{01}\delta_0^A \delta_1^B+h_{01}\delta_0^B \delta_1^A+ \frac1{f^2} h_{02} \delta_0^A \delta_2^B + \frac1{f^2} h_{02} \delta_0^B \delta_2^A \right) \nonumber \\
& = \left( \delta_1^A - \frac{h_{01}}{h_{00}} \delta_0^A  \right) \left( \delta_1^B - \frac{h_{01}}{h_{00}} \delta_0^B  \right) + 
\frac1{f^2} \left( \delta_2^A - \frac{h_{02}}{h_{00}} \delta_0^A  \right) \left( \delta_2^B - \frac{h_{02}}{h_{00}} \delta_0^B  \right)
=L_1^A L_1^B + L_2^A L_2^B \,. \label{tensork}
\end{align}
Notice that $L_1$ and $L_2$ are obtained by subtracting from $\partial_{\lambda^1}$ and $(1/f) \partial_{\lambda^2}$ (that is, $\delta_1^A$ and $(1/f) \delta_2^A$) their orthogonal projections along $\partial_{\tau}$ (that is, $\delta_0^A$). If we use the derivative of the embedding $X\colon \mathbb{R}\times S\to M$ to identify tangent spaces to $S$ with hyperplanes in the worldtube orthogonal to the four-velocity $U^A$, then $k^{AB}$ coincides with the contravariant relaxed metric in those hyperplanes.  Note that the covariant relaxed metric is {\em not} $k_{AB}=h_{AC}h_{BD}k^{CD}$, but is instead the inverse to the restriction of $k^{AB}$ to the orthogonal hyperplanes; moreover, $L_1$ and $L_ 2$ are orthonormal for the relaxed metric, but {\em not} for the metric $h_{AB}$. Note also that
\begin{equation}
h_{AB} k^{AB} = h_{11}+\frac{h_{22}}{f^2}-\frac{1}{h_{00}}\left(h_{01}^2+\frac{h_{02}^2}{f^2}\right) = s^2,
\end{equation}
as one would expect, and so the trace part of $k^{AB}$ is
\begin{equation}
\frac{s^2}2 \left(U^A U^B + h^{AB}\right) = \frac{s^2}2 \left(-\frac1{h_{00}}\delta_0^A \delta_0^B + h^{AB}\right).
\end{equation}

By analogy with the energy-momentum tensor in general relativity, we define the membrane's energy-momentum tensor $T^{AB}$ by the relation
\begin{equation}\label{EMTensor}
\delta \mathcal{L}=-\frac{1}{2} \sqrt{-h} \,T^{A B} \delta h_{A B}\,.
\end{equation}
Then, the membrane's energy-momentum tensor is given by
\begin{align} \label{TAB}
T^{AB}= 2n^2 F^{(1,0)} U^A U^B+\left(2 n^2 F^{(1,0)}-F\right) h^{A B} - 2F^{(0,1)} k^{AB}\,.
\end{align}
Note that in the case of an isotropic deformation $k^{A B}$ reduces to its trace part. Therefore, the membrane's energy density $\rho$ and isotropic pressure $p$ are given by
\begin{equation} \label{rhoandp}
\rho=F\,, \qquad \qquad p= 2 n^2 F^{(1,0)}-s^2F^{(0,1)}-F\,.
\end{equation}
To derive the equations of motion, we note that
\begin{equation}
\delta h_{AB} = \partial_\alpha g_{\mu\nu} \delta X^\alpha \partial_A X^\mu \partial_B X^\nu + 2 g_{\mu\nu} \partial_A X^\mu \partial_B \delta X^\nu \, ,
\end{equation}
and so
\begin{equation}
- 2 \delta \cL = \left[ \sqrt{-h} \, T^{AB} \partial_\alpha g_{\mu\nu} \partial_A X^\mu \partial_B X^\nu - \partial_B \left( 2 \sqrt{-h} \, T^{AB} g_{\mu\alpha} \partial_A X^\mu \right)\right] \delta X^\alpha + \partial_B \left( 2 \sqrt{-h} \, T^{AB} g_{\mu\alpha} \partial_A X^\mu \delta X^\alpha \right) .
\end{equation}
Discarding the total divergence in Hamilton's principle
\begin{equation}
\delta \int_{\bbR \times I} \delta\cL(X,\partial X) \, d\tau d\lambda = 0 \, ,
\end{equation}
we obtain the equations of motion in the form
\begin{equation} \label{motion}
\frac1{\sqrt{-h}}\partial_B \left( \sqrt{-h} \, T^{AB}\partial_A X^\alpha \right) + T^{AB} \Gamma^\alpha_{\mu\nu} \partial_A X^\mu \partial_B X^\nu = 0\,\,.
\end{equation}
These equations are formally the same as the equations of motion for elastic strings, and are closely related to the harmonic map/wave map/nonlinear sigma model equations (see for instance \cite{Jost02, Tao06, Rendall08, NQV18}). Their well-posedness has been studied in the particular case of the Nambu-Goto membrane in Minkowski spacetime \cite{Lind04, WW22} (see also \cite{Wong11}).

\subsection{Adapted coordinates} 

To better understand the equations of motion, we extend the local coordinates $(x^A)=(\tau,\lambda^1, \lambda^2)$ on the worldtube $\Sigma$ to a local coordinate system $(x^A,x^i)$ defined on a neighborhood of $\Sigma$ in the following way: we choose an orthonormal frame $\{E_3, \ldots, E_n\}$ for the normal bundle of the worldtube, and parameterize by $(x^A, x^i)$ the point $\exp_p(x^i E_i)$, where $\exp_p$ is the geodesic exponential map and $p \in \Sigma$ is the point with coordinates $(x^A)$. The worldtube is given in these coordinates by $x^i=0$, and the spacetime metric by
\begin{equation}
g = g_{AB} dx^A dx^B + 2 g_{Ai} dx^A dx^i + g_{ij} dx^i dx^j \, .
\end{equation}
Note that on $\Sigma$ we have
\begin{equation}
g_{|_{\Sigma}} = h_{AB} dx^A dx^B + \delta_{ij} dx^i dx^j \, .
\end{equation}
The tensor
\begin{equation}
K^i_{AB} = \frac12\partial_i g_{AB} \, ,
\end{equation}
defined on $\Sigma$, is called the {\em extrinsic curvature} (or {\em second fundamental form}) of $\Sigma$ in the direction of $E_i$. It easily seen that on $\Sigma$
\begin{equation}
\Gamma^i_{AB} = -K^i_{AB}
\end{equation}
and
\begin{equation}
\Gamma^C_{AB} = \overline{\Gamma}^C_{AB} \, ,
\end{equation}
where $\overline{\Gamma}^C_{AB}$ are the Christoffel symbols for the Levi-Civita connection $\overline{\nabla}$ of $h_{AB}$. In this coordinate system, the embedding is simply given by $(X^A,X^i)=(x^A,0)$, and so we can write the first three components of equation~\eqref{motion} as
\begin{equation} \label{conservation}
\frac1{\sqrt{-h}}\partial_B \left( \sqrt{-h}T^{BC} \right) + T^{AB} \overline{\Gamma}^C_{AB} = 0 \, ,
\end{equation}
and the last $n-2$ as the so-called {\em generalized sail equations}
\begin{equation} \label{constraints}
T^{AB} K^i_{AB} = 0 \,\,.
\end{equation}
Using the well-known formula
\begin{equation}
\partial_B \log \sqrt{-h} = \overline{\Gamma}^A_{BA} \, ,
\end{equation}
equation~\eqref{conservation} is easily seen to be equivalent to
\begin{equation} \label{conservation2}
\overline{\nabla}_B T^{BC} = 0 \,\,.
\end{equation}
This justifies the choice of $T^{AB}$ as the membrane's energy-momentum tensor. 

The equations of motion of the membrane can then be understood as constraints of the geometry of the worldtube, given by \eqref{constraints}, plus conservation of energy-momentum, given by \eqref{conservation2}.\footnote{These equations occur for other extended objects such as branes \cite{Carter92, Carter01, Carter11} and blackfolds \cite{EHNO10}.} For Nambu-Goto membranes, for instance, where $T_{AB}$ is proportional to $h_{AB}$, the constraints are the condition that the worldtube is a minimal surface, and the conservation equation is automatically satisfied.

Note that
\begin{align}
-U_B\overline{\nabla}_A T^{A B}&=\overline{\nabla}_A\left(2n^2 F^{(1,0)}U^A\right)-U^A\overline{\nabla}_A\left(2 n^2 F^{(1,0)}-F\right)+2F^{(0,1)}U_B\overline{\nabla}_A k^{A B}\nonumber \\
&=2n^2 F^{(1,0)}\overline{\nabla}_AU^A+F^{(1,0)}U^A\overline{\nabla}_An^2+F^{(0,1)}U^A\overline{\nabla}_As^2+2F^{(0,1)}U_B\overline{\nabla}_A k^{A B}\nonumber \\
&=2 n F^{(1,0)} \overline{\nabla}_A\left(n\, U^A\right)+F^{(0,1)}\left(U^A\overline{\nabla}_As^2+2U_B\overline{\nabla}_A k^{A B}\right)\nonumber \\
&=F^{(0,1)}\left(U^A\overline{\nabla}_As^2-2k^{A B} \overline{\nabla}_A U_B\right) \,,
\end{align}
where we used the orthogonality relations
\begin{equation}
U_B \overline{\nabla}_A U^B = 0 \, , \qquad k^{AB} U_B = 0 \, ,
\end{equation}
and the \textit{conservation of number of particles},
\begin{align}
\overline{\nabla}_U\left(n U^A\right)=\frac{1}{\sqrt{-h}}\partial_A (\sqrt{-h}\, n\, U^A)=\frac{1}{\sqrt{-h}}\partial_A \left(f \, \delta^A_0\right)=0\,.
\end{align}
Note that
\begin{equation}
h_{AB} \mathcal{L}_U k^{AB} = h_{AB} \left(\overline{\nabla}_U k^{AB} - k^{CB} \overline{\nabla}_C U^A - k^{AC} \overline{\nabla}_C U^B\right) = \overline{\nabla}_Us^2- 2 k^{A B} \overline{\nabla}_A U_B \,.
\end{equation}
Using the fact that
\begin{align}
&\mathcal{L}_U L_1 = \left[U, L_1\right] = \left[\tfrac{1}{\alpha}\partial_\tau, \partial_{\lambda^1} - \tfrac{\beta_1}{\alpha}\partial_\tau \right] = \alpha g_1 \partial_\tau = g_1 U \,, \\
&\mathcal{L}_U L_2 = \left[U, L_2\right] =  \left[\tfrac{1}{\alpha}\partial_\tau, \tfrac{1}{f}\partial_{\lambda^2} - \tfrac{\beta_2}{\alpha}\partial_\tau \right] = \alpha g_2 \partial_\tau = g_2 U\,,
\end{align}
for appropriate functions $g_1$ and $g_2$, we have
\begin{equation}
\mathcal{L}_U k^{AB} = \mathcal{L}_U \left(L_1^{A}L_1^{B} + L_2^{A}L_2^{B} \right) = g_1 \left(U^{A}L_1^{B} + L_1^{A}U^{B} \right) + g_2 \left(U^{A}L_2^{B} + L_2^{A}U^{B} \right) \,,
\end{equation}
and so
\begin{equation}
h_{AB} \mathcal{L}_U k^{AB} = 0
\end{equation}
(since both~$L_1$ and~$L_2$ are orthogonal to~$U$).
We conclude that the component of the conservation equations along $U^A$ is always identically satisfied,
 \begin{align}
 -U_B\overline{\nabla}_A T^{A B}=0\,.
 \end{align}

\subsection{Conserved quantities} 

If $(M,g)$ admits a Killing vector field $\xi$,
\begin{equation}
\nabla_{(\mu} \xi_{\nu)} = 0\,,
\end{equation}
then in the coordinates above we have
\begin{equation}
\partial_{(A} \xi_{B)} + \overline{\Gamma}^C_{AB} \xi_{C} + \Gamma^i_{AB} \xi_i = 0 \Leftrightarrow \overline\nabla_{(A} \xi_{B)} 
= K^i_{AB} \xi_i\,,
\end{equation}
that is, the projection of $\xi$ on $T\Sigma$ is not, in general, a Killing vector field of $h_{AB}$. Nevertheless,
\begin{equation}
\overline\nabla_{A} (T^{AB} \xi_{B}) =  T^{AB} K^i_{AB} \xi_i = 0\,,
\end{equation}
in view of \eqref{constraints}, that is, the vector field
\begin{equation}
j^A = T^{AB} \xi_{B}
\end{equation}
is divergenceless on $\Sigma$. As a consequence, the quantity
\begin{equation}
E^\xi = \int_{\{\tau=\text{constant}\}} j^A \nu_A \sqrt{h_{11}h_{22}-h_{12}^2} \,\, d\lambda^1 d\lambda^2 
\end{equation}
is conserved, where 
\begin{equation}
\nu_A = \frac{\delta^0_A}{\sqrt{-h^{00}}} = \frac{\sqrt{-h}}{\sqrt{h_{11}h_{22}-h_{12}^2}} \, \delta^0_A 
\end{equation}
is the past-pointing normal to the spacelike surface $\{\tau=\text{constant}\}$. In other words, we have the conserved quantity
\begin{equation}\label{conserved_quantity}
E^\xi = \int_{\{\tau=\text{constant}\}} j^0 \sqrt{-h} \, d\lambda^1 d\lambda^2\,.
\end{equation}

\subsection{Speeds of sound}

The speeds of local perturbations traveling on a membrane can be obtained by linearizing the equations of motion about a (possibly isotropically stretched) stationary membrane in Minkowski spacetime lying on the $z=0$ plane. This corresponds to taking terms up to quadratic order in the Lagrangian obtained from the embedding\footnote{Without loss of generality, since the membrane is isotropically stretched, we consider perturbations travelling along the $x$-direction.}
\begin{equation}
\begin{cases}
t(\tau, \lambda^1,\lambda^2) = \tau \\
x(\tau, \lambda^1, \lambda^2) = n_0^{-1/2} \lambda^1 + \delta x(\tau, \lambda^1) \\
y(\tau, \lambda^1,\lambda^2) =n_0^{-1/2} \lambda^2+ \delta y(\tau, \lambda^1) \\
z(\tau, \lambda^1,\lambda^2) = \delta z(\tau, \lambda^1)
\end{cases} .
\end{equation}
To compute $h_{AB}$ to quadratic order it suffices to use the approximation
\begin{equation}
(h_{AB}) = 
\left(
\begin{matrix}
- 1 + \delta \dot{x}^2 + \delta \dot{y}^2 + \delta \dot{z}^2 & n_0^{-1/2} \delta \dot{x}+ \delta \dot{x} \delta x' +\delta \dot{y} \delta y'+ \delta \dot{z} \delta z'& n_0^{-1/2} \delta \dot{y} \\
n_0^{-1/2} \delta \dot{x}+ \delta \dot{x} \delta x' +\delta \dot{y} \delta y'+ \delta \dot{z} \delta z' & \left(n_0^{-1/2}+\delta x'\right)^2+(\delta y')^2+(\delta z')^2 & n_0^{-1/2}\delta y'  \\
n_0^{-1/2} \delta \dot{y} & n_0^{-1/2}\delta y' & n_0^{-1}
\end{matrix}
\right),
\end{equation}
where $\,\,\dot{} \equiv \frac{\partial}{\partial \tau}$ and $\,\,'\equiv \frac{\partial}{\partial \lambda^1}$, so that
\begin{equation}
n^2=n_0^2\left(1-2 \sqrt{n_0}\, \delta x'-\delta \dot{x}^2-\delta \dot{y}^2-n_0\left(\delta z'\right)^2+3 n_0\left(\delta x'\right)^2\right)
\end{equation}
and
\begin{equation}
s^2=\frac{2}{n_0}\left(1+\sqrt{n_0}\,\delta x'+\delta \dot{x}^2+\delta \dot{y}^2+\frac{n_0}{2} \left[\left(\delta x'\right)^2+\left(\delta y'\right)^2+\left(\delta z'\right)^2\right]\right)\,.
\end{equation}
Note that for an isotropically stretched membrane we have
\begin{equation}
s_0^2=\frac{2}{n_0}\,.
\end{equation}
Using the Taylor formula to second order,
\begin{align}
F(n^2,s^2)=&F\left(n_0^2,\frac{2}{n_0}\right)+F^{(1,0)}\left(n^2-n_0^2\right)+F^{(0,1)}\left(s^2-\frac{2}{n_0}\right)\nonumber \\
&+\frac{1}{2}F^{(2,0)}\left(n^2-n_0^2\right)^2+\frac{1}{2}F^{(0,2)}\left(s^2-\frac{2}{n_0}\right)^2+ F^{(1,1)}\left(n^2-n_0^2\right)\left(s^2-\frac{2}{n_0}\right)\,,
\end{align}
and discarding constants and total divergences, we finally obtain
\begin{align}
\mathcal{L}=& \,\, F\left(n^2,s^2\right)\sqrt{-h}=\frac{F}{2 n_0}\left[\left(\frac{n_0 F+ 2 F^{(0,1)}-2n_0^3 F^{(1,0)}}{F}\right)\left(\delta z'\right)^2-\delta \dot{z}^2\right] \nonumber \\
&+\frac{n_0^3 F^{(1,0)}-F^{(0,1)}}{n_0^{2}}\left[\left(\frac{2n_0^6 F^{(2,0)}+2F^{(0,2)}-4n_0^3F^{(1,1)}+n_0^4F^{(1,0)}+3n_0 F^{(0,1)}}{n_0^3 F^{(1,0)}-F^{(0,1)}}\right)\left(\delta x'\right)^2-\delta \dot{x}^2\right] \nonumber \\
&+\frac{n_0^3 F^{(1,0)}-F^{(0,1)}}{n_0^2}\left[\left(\frac{n_0 F^{(0,1)}}{n_0^3F^{(1,0)}-F^{(0,1)}}\right)\left(\delta y'\right)^2-\delta \dot{y}^2\right]\,.
\end{align}
Therefore, the perturbations $\delta z$, $\delta x$ and $\delta y$ satisfy the wave equation in the coordinates $\left(\tau,\lambda^1\right)$ with wave speeds, respectively,
\begin{equation}
c'_T=\sqrt{n_0}\,\sqrt{\frac{ F+ 2 n_0^{-1} F^{(0,1)}-2n_0^2 F^{(1,0)}}{F }}\,,
\end{equation}
\begin{equation}
c'_L=\sqrt{n_0} \,\sqrt{\frac{2n_0^5 F^{(2,0)}+2n_0^{-1}F^{(0,2)}-4n_0^2F^{(1,1)}+n_0^3F^{(1,0)}+3 F^{(0,1)}}{n_0^3 F^{(1,0)}-F^{(0,1)}}}\,,
\end{equation}
and
\begin{equation}
c'_{TT}=\sqrt{n_0}\, \sqrt{\frac{ F^{(0,1)}}{n_0^3F^{(1,0)}-F^{(0,1)}}}\,.
\end{equation}
Since $\lambda^1=\sqrt{n_0} \,x$ for the isotropically stretched membrane, one sees that the physical speed of sound for transverse waves orthogonal to the plane of the membrane is
\begin{equation}
c_T=\sqrt{\frac{ F+ 2 n_0^{-1} F^{(0,1)}-2n_0^2 F^{(1,0)}}{F }}=\sqrt{-\frac{p}{\rho}}\,,
\end{equation}
whereas the physical speeds of sound for longitudinal and transverse waves in the plane of the membrane are, respectively,
\begin{equation}\label{velL}
c_L=\sqrt{\frac{2n_0^5 F^{(2,0)}+2n_0^{-1}F^{(0,2)}-4n_0^2F^{(1,1)}+n_0^3F^{(1,0)}+3 F^{(0,1)}}{n_0^3 F^{(1,0)}-F^{(0,1)}}}
\end{equation}
and
\begin{equation}\label{velTin}
c_{TT}= \sqrt{\frac{ F^{(0,1)}}{n_0^3F^{(1,0)}-F^{(0,1)}}}\,.
\end{equation}
Notice that the speed of sound for transverse waves oscillating orthogonally to the membrane's plane is determined by the membrane's tension, just as for strings \cite{NQV18}. In particular, these waves do not exist if the membrane is not under tension, and in fact their speed is imaginary when the membrane is compressed, signaling a buckling instability. The speeds of sound for longitudinal waves and transverse waves oscillating in the membrane's plane, on the other hand, are nonzero even when the membrane is relaxed, and can be seen as a measure of the membrane's longitudinal and transverse rigidity. It is interesting to note that, since the isotropic pressure on an isotropic state is given by
\begin{equation}
p= 2 n_0^2 F^{(1,0)}\left(n_0^2, \frac2{n_0}\right)-\frac2{n_0}F^{(0,1)}\left(n_0^2, \frac2{n_0}\right)-F\left(n_0^2, \frac2{n_0}\right)\,,
\end{equation}
we have
\begin{equation}
\frac{dp}{dn_0}= 4n_0^3 F^{(2,0)}+4n_0^{-3}F^{(0,2)}-8F^{(1,1)}+2n_0F^{(1,0)}+4 n_0^{-2} F^{(0,1)}
\end{equation}
and
\begin{equation}
\frac{d\rho}{dn_0}= 2n_0F^{(1,0)} - 2n_0^{-2}F^{(0,1)}\,,
\end{equation}
so that
\begin{equation}
\frac{dp}{d\rho} = c_L^2 - c_{TT}^2 \, .
\end{equation}
\subsection{Elastic coefficients}

Unlike the elastic strings studied in \cite{NQV18}, elastic membranes have intrinsic geometry. Consequently, their linear elasticity is characterized by the two-dimensional versions of the elastic constants in the usual theory of linear elasticity, which we now discuss.

\subsubsection{Poisson ratio}\label{ssPoisson}

The Poisson ratio measures how much the membrane stretches in a given direction when compressed in the orthogonal direction. To determine it, we compute minus the ratio between the infinitesimal increment in the expansion factor along the $xx$-axis and the infinitesimal increment in the expansion factor along the $yy$-axis, when the membrane is compressed along the $yy$-axis and unconstrained along the $xx$-axis. Denoting these by $\alpha$ and $\beta$, respectively, we consider the embedding
\begin{equation}\label{embedding}
\begin{cases}
t(\tau, \lambda^1,\lambda^2) = \tau \\
x(\tau, \lambda^1,\lambda^2) = \left(1+\alpha\right) \lambda^1 \\
y(\tau, \lambda^1, \lambda^2) =\left(1+\beta\right) \lambda^2 \\
z(\tau, \lambda^1,\lambda^2) = 0
\end{cases} .
\end{equation}
The induced metric is then given by
\begin{equation}
(h_{AB}) = 
\left(
\begin{matrix}
- 1  & 0 & 0 \\
0 & (1+\alpha)^2 & 0  \\
0 & 0 & (1+\beta)^2
\end{matrix}
\right) .
\end{equation}
Having the $xx$-axis unconstrained corresponds to the condition $T^{xx}=0$ and so, by~\eqref{deltaL} and~\eqref{EMTensor}, this can be written, to first order in $(\alpha,\beta)$, as
\begin{equation}\label{Poisson}
\begin{array}{lll}
&-F+2\left(F^{(1,0)}-F^{(0,1)}\right)\\[8pt]
&-2\left(-F+3F^{(1,0)}+F^{(0,1)}-4F^{(1,1)}+2F^{(2,0)}+2F^{(0,2)}\right)\alpha\\[8pt]
&-2\left(F^{(1,0)}+F^{(0,1)}-4F^{(1,1)}+2F^{(2,0)}+2F^{(0,2)}\right)\beta=0 \, ,
\end{array}
\end{equation}
which implies, from the zeroth order term, that
\begin{equation}\label{equil}
-F+2\left(F^{(1,0)}-F^{(0,1)}\right)=0 \, .
\end{equation}
Note that this is just the condition that must be verified for $\alpha,\beta=0$ to be the (unstressed) relaxed configuration, and we will use it from here on to remove any explicit dependencies in $F$. Now, using~\eqref{Poisson} to first order, we obtain the Poisson ratio
\begin{equation}\label{Poissonformula}
\nu=-\frac{\alpha}{\beta}=\frac{\tilde{F}}{2F^{(0,1)}+\tilde{F}} \, ,
\end{equation}
where
\begin{equation}\label{tildeF}
\tilde{F}\defeq F^{(1,0)}+F^{(0,1)}-4F^{(1,1)}+2F^{(2,0)}+2F^{(0,2)} \, .
\end{equation}
It can easily be checked from~\eqref{velL} and~\eqref{velTin} (with $n_0=1$) that the following relation holds:
\begin{equation}
\frac{c_{TT}^2}{c_L^2}=\frac{1-\nu}{2} \, .
\end{equation}

\subsubsection{Bulk modulus}
The bulk modulus measures the pressure required to isotropically compress an element of the membrane by a small amount. It is given by minus the ratio between the infinitesimal isotropic pressure and the infinitesimal fractional increment in area. Hence we set $\alpha=\beta$ in~\eqref{embedding}, and so the fractional increment in volume becomes
\begin{equation}
(1+\alpha)^2=1+2\alpha \, .
\end{equation}
On the other, we obtain
\begin{equation}
T^{xx}=T^{yy}=-4\left(F^{(1,0)}+2F^{(0,1)}-4F^{(1,1)}+2F^{(2,0)}+2F^{(0,2)}\right)\alpha \, .
\end{equation}
Therefore, the bulk modulus is given by
\begin{equation}\label{Kformula}
K=-\frac{T^{xx}}{2\alpha}=2F^{(1,0)}+4F^{(0,1)}-8F^{(1,1)}+4F^{(2,0)}+4F^{(0,2)} = 2F^{(0,1)} + 2\tilde{F} \, ,
\end{equation}
which is also related with the speeds of sound in the relaxed configuration by
\begin{equation}\label{BulkVelRel}
c_L^2-c_{TT}^2=\frac{K}{\rho_0}
\end{equation}
(where $\rho_0$ is the density of the relaxed configuration). Thermodynamic stability considerations usually require that $K>0$ for physical materials.

\subsubsection{Shear modulus}
The shear modulus measures the pressure required to shear an element of the membrane by a small angle. It is given by minus the ratio between the infinitesimal shear pressure and the infinitesimal shear angle. A shear deformation by an infinitesimal angle $\alpha$ corresponds to the embedding
\begin{equation}\label{embedding1}
\begin{cases}
t(\tau, \lambda^1,\lambda^2) = \tau \\
x(\tau, \lambda^1,\lambda^2) = \lambda^1 -\alpha \lambda^2 \\
y(\tau, \lambda^1, \lambda^2) =\lambda^2 \\
z(\tau, \lambda^1,\lambda^2) = 0
\end{cases} \, ,
\end{equation}
corresponding to a shear deformation in the $xy$-plane by an infinitesimal angle $\alpha$. The induced metric is then
\begin{equation}
(h_{AB}) = 
\left(
\begin{matrix}
- 1  & 0 & 0 \\
0 & 1 & -\alpha  \\
0 & -\alpha & 1+\alpha^2
\end{matrix}
\right),
\end{equation}
and so we get
\begin{equation} \label{Gformula}
G=-\frac{T^{xy}}{\alpha}=2F^{(0,1)}
\end{equation}
for the shear modulus. It it be related with the tangential transverse speed of sound~\eqref{velTin} in the relaxed configuration by
\begin{equation}
c_{TT}^2=\frac{G}{\rho} \, .
\end{equation}
Combining this with~\eqref{BulkVelRel} allows us to write
\begin{equation}
c_L^2=\frac{K+G}{\rho}
\end{equation}
for the longitudinal speed of sound.

\subsubsection{Young modulus}
The Young modulus measures the pressure required to compress an element of the membrane by a small amount along a given direction. To determine it, we compute minus the ratio between the infinitesimal pressure along the $yy$-axis, $T^{yy}$, and the infinitesimal increment in the expansion factor along the $yy$-axis, when the membrane is compressed along the $yy$-axis and unconstrained along the $xx$-axis. Therefore, we again consider the setup of subsection~\ref{ssPoisson} and and note that the Young modulus is
\begin{equation}
E = - \frac{T^{yy}}{\beta} \, .
\end{equation}
Now equations~\eqref{deltaL}, \eqref{EMTensor} and \eqref{equil} imply that, to first order in $\alpha$ and $\beta$,
\begin{equation}\label{Young}
T^{yy} = - 2 \tilde{F}\alpha -2\left(2F^{(0,1)}+\tilde{F}\right)\beta \, ,
\end{equation}
and so
\begin{equation}
E = 2\tilde{F} \frac{\alpha}{\beta} + 2\left(2F^{(0,1)}+\tilde{F}\right) \, ,
\end{equation}
or, using \eqref{Poissonformula},
\begin{equation}
E = \frac{8F^{(0,1)}\left(F^{(0,1)}+\tilde{F}\right)}{2F^{(0,1)}+\tilde{F}} \, .
\end{equation}
From \eqref{Poissonformula}, \eqref{Kformula} and \eqref{Gformula} we have
\begin{equation}
E = 2G(1+\nu) = 2(1 - \nu) K \, .
\end{equation}
%
\section{Rigidly rotating rigid disk}
\label{section2}
In this section, we give an example of a simple $1$-parameter family of elastic laws, which we dub the {\em rigid membrane}, since its longitudinal speed of sound is equal to the speed of light. We determine the transverse speeds of sound, and compute the coefficients of linear elasticity with respect to the relaxed configuration. Then we use this elastic law to produce, for the first time, explicit examples of fully relativistic rigidly rotating disks, which have been widely discussed in the context of Ehrenfest's paradox \cite{Ehrenfest09, Gron04} (see also \cite{Brotas68, McCrea71, NQV18} for similar discussions in the simpler context of rotating elastic rings).
\subsection{Rigid membrane}
We consider the elastic law
\begin{equation}\label{rigid_law}
F(n^2,s^2) = \frac{\rho_0}{2}\left[\left( 1 - \epsilon \right) \left(1 + n^2 \right) + \epsilon n^2s^2\right] \, ,
\end{equation}
which has the special property that the longitudinal speed of sound in any isotropic state is always equal to the speed of light:
\begin{equation}
c_L^2 = 1 \, .
\end{equation}
Accordingly, we dub this elastic law the {\em rigid membrane}. Apart from the overall dimensionfull factor $\rho_0$, which does not change the equations of motion, this elastic law depends only on the adimensional parameter $\epsilon$, giving us a $1$-parameter family of materials.

The remaining sound speeds satisfy
\begin{equation}
c_T^2 = \frac{\left( 1 - \epsilon \right) \left(1 - n^2\right)}{\left( 1 - \epsilon \right) \left(1 + n^2 \right) + 2 \epsilon n} \, ,
\end{equation}
\begin{equation}
c_{TT}^2 = \frac{\epsilon}{n \left( 1 - \epsilon + \epsilon s^2 \right) - \epsilon} \, ,
\end{equation}
and the isotropic pressure is
\begin{equation}
p = \frac{\rho_0}2\left( 1 - \epsilon \right) \left(n^2 - 1\right) \, .
\end{equation}
In particular, in the relaxed state (where $n=1$ and $s^2=2$), we have
\begin{align}
& p=0 \, , \\
& c_T^2 = 0 \, , \\
& c_{TT}^2 = \epsilon \, . \label{epsilon}
\end{align}
Therefore the parameter $\epsilon$ can be regarded as a measure of the transverse rigidity of the material, which increases with increasing $\epsilon$. Note that \eqref{epsilon} implies that $0 < \epsilon < 1$ for physically reasonable materials.

The Poisson ratio, bulk modulus, shear modulus and Young modulus are, respectively,
\begin{align}
& \nu = 1 - 2 \epsilon \, , \\
& K = \rho_0(1 -  \epsilon ) \, , \\
& G = \rho_0 \epsilon \, , \\
& E = 4 \rho_0 \epsilon \left( 1 - \epsilon \right) \, .
\end{align}
Note that the bulk modulus is positive in the range $0 < \epsilon < 1$, but the Poisson ratio becomes negative for~$\epsilon> 1/2$. It is easily checked that one indeed has the standard relations
\begin{align}
& \frac{1-\nu}2  = c_{TT}^2 \, , \\
& \frac{K}{\rho_0} = c_L^2 - c_{TT}^2 \, , \\
& \frac{G}{\rho_0} = c_{TT}^2 \, , \\
& E = 2G(1+\nu) = 2(1 - \nu) K \, .
\end{align}
\subsection{Rotating disk}
\begin{figure}
	\centering
	\begin{subfigure}{0.45\textwidth}
		\includegraphics[height = 6.2 cm]{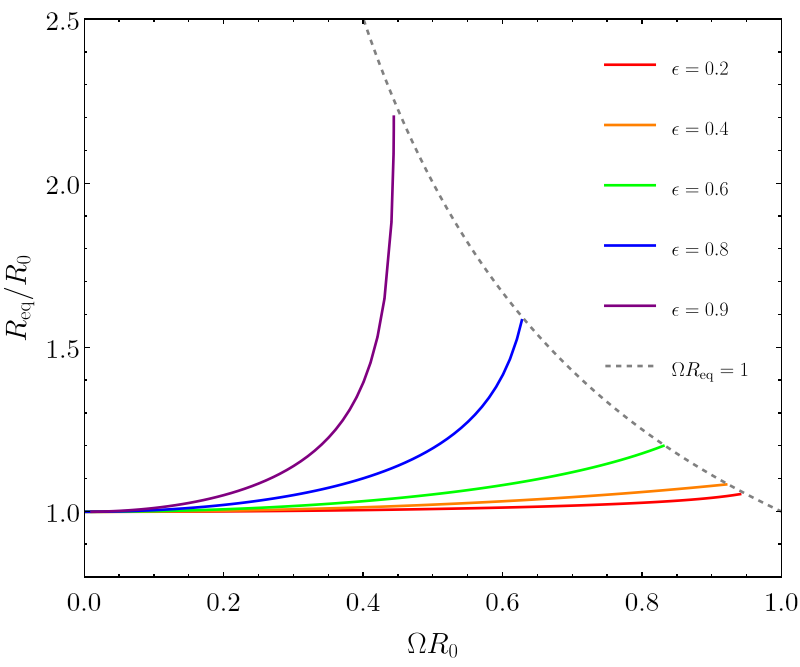}	
	\end{subfigure}
	\hfill
	\begin{subfigure}{0.45\textwidth}
		\includegraphics[height = 6 cm]{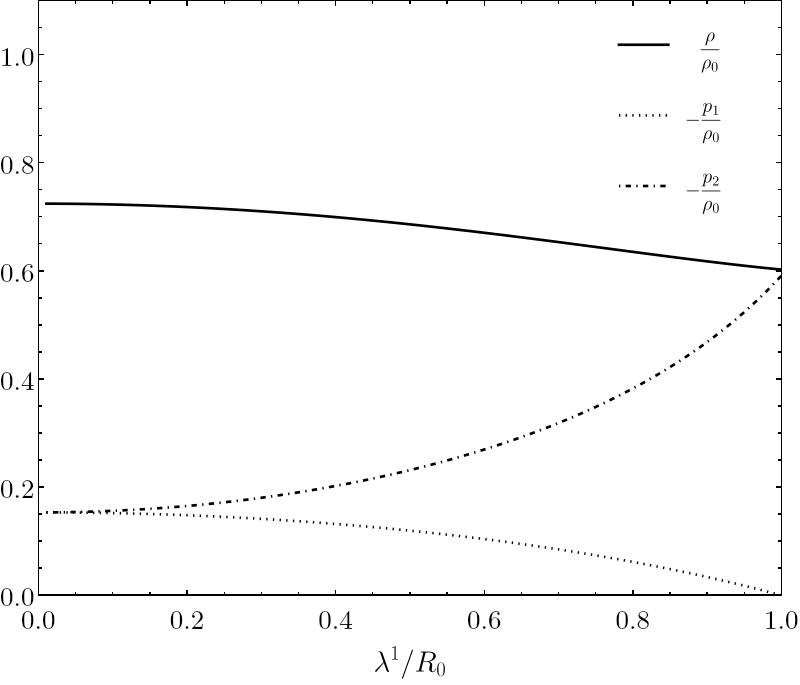}
	\end{subfigure}
	\hfill	
	\caption{\textsc{Left:} Rigid disk's equilibrium radii as function of angular velocity, for different rigid membranes~\eqref{rigid_law}. The membrane's edge never becomes superluminal for these elastic laws. \textsc{Right:} Energy density and principal pressures as a function of~$\lambda^1$ for~$\epsilon=0.4$ and near the maximal angular velocity~$\Omega\approx0.92$. The dominant energy condition is saturated at the disk's edge, due to its nearly luminal velocity.} \label{fig:Req}
\end{figure}

Let us start by writing the Minkowski metric in spherical coordinates:
\begin{equation}\label{M_metric}
ds^2 = - dt^2 + dr^2 + r^2 \left(d \theta^2 + \sin^2 \theta d\varphi^2\right) \, .
\end{equation}
Consider a rigid disk of radius $R_0$, corresponding to the relaxed metric
\begin{equation}
k = \left(d\lambda^1\right)^2 + \left(\lambda^1\right)^2\left(d\lambda^2\right)^2 \, ,
\end{equation}
with $(\lambda^1, \lambda^2) \in \left(0, R_0\right) \times \left(0,2\pi \right)$. As discussed in Section~\ref{section1}, the motion of this membrane in the Minkowski geometry is described by an embedding, which we take to be of the form
\begin{equation}\label{embedding_RigidRot}
\begin{cases}
t(\tau, \lambda^1,\lambda^2) = \tau \\
r(\tau, \lambda^1,\lambda^2) = R(\lambda^1)  \\
\theta(\tau, \lambda^1, \lambda^2) =\frac{\pi}{2} \\
\varphi(\tau, \lambda^1,\lambda^2) = \Omega \tau+\lambda^ 2
\end{cases} ,
\end{equation}
corresponding to a rigid rotation motion with constant angular velocity $\Omega$. The metric induced on the worldtube by the embedding in flat space is
\begin{equation}
(h_{AB}) = 
\left(
\begin{matrix}
- 1+ R^2\Omega^2  & 0 & R^2 \Omega \\
0 & R'^2 &   0\\
R^2 \Omega & 0 & R^2
\end{matrix}
\right).
\end{equation}
and so
\begin{equation} \label{n^2}
n^2 = \frac{h_{00}f^2}{h} =  \frac{\left(1-R^2\Omega^2\right)\left(\lambda^1\right)^2}{R^2 R'^2}
\end{equation}
and
\begin{equation} \label{s^2}
s^2 = h_{11} + \frac{h_{22}}{f^2} - \frac1{h_{00}} \left( h_{01}^2 + \frac{h_{02}^2}{f^2} \right) = R'^2 + \frac{R^2}{\left(1-R^2\Omega^2\right)\left(\lambda^1\right)^2} \, .
\end{equation}
Moreover, we have
\begin{equation}
(h^{AB}) = 
\left(
\begin{matrix}
-1  & 0 & \Omega \\
0 & \frac1{R'^2} &   0\\
\Omega & 0 & \frac{1- R^2\Omega^2}{R^2}
\end{matrix}
\right)
\end{equation}
and
\begin{equation}
(k^{AB}) = 
\left(
\begin{matrix}
\frac{R^4\Omega^2}{\left(1-R^2\Omega^2\right)^2\left(\lambda^1\right)^2}  & 0 & \frac{R^2\Omega}{\left(1-R^2\Omega^2\right)\left(\lambda^1\right)^2} \\
0 & 1 &   0\\
\frac{R^2\Omega}{\left(1-R^2\Omega^2\right)\left(\lambda^1\right)^2} & 0 & \frac1{\left(\lambda^1\right)^2}
\end{matrix}
\right) \, ,
\end{equation}
so that, substituting in \eqref{TAB} and \eqref{motion}, the equilibrium condition for the rigid membrane given by~\eqref{rigid_law} is given by
\begin{align}
	&\left[(1-\epsilon)\big[1-(\Omega R) ^2\big](\tfrac{\lambda^1}{R})^2+\epsilon\right]R''+ \epsilon\tfrac{\lambda^1}{R} \big[1-(\Omega R)^2\big] \frac{R'^3}{R}\nonumber \\
	&\qquad-\left(\left[\epsilon-(\lambda^1 \Omega)^2 (1-\epsilon)\right]- (1-\epsilon)(\tfrac{\lambda^1}{R})^2\right)\frac{R'^2}{R} - (1-\epsilon) \lambda^1\left[1-(\Omega R)^2\right]\frac{R'}{R^2}=0 \, .
\end{align}
%
This highly nonlinear equation can be solved numerically imposing the boundary conditions
\begin{equation}
R(0)=0, \qquad T^{11}(R_0)=0 \, ,
\end{equation}
i.e., by requiring that the disk's geometry is regular at the origin and the radial pressure vanishes at the boundary, where, from \eqref{TAB}, we have
\begin{align}
	T^{11}=\frac{(\lambda^1)^2\left(1-\epsilon-\epsilon R'^2)\right)-R^2 \left[\epsilon-(1-\epsilon)(\lambda^1 \Omega)^2-\left(1-\epsilon-\epsilon\,(\lambda^1 \Omega)^2\right)R'^2\right]}{2 R^2 R'^4} \, .
\end{align}
The left panel of Figure~\ref{fig:Req} shows the equilibrium radius of the rotating disk as function of its angular velocity, for different values of~$0<\epsilon<1$. In this range, we find that the disk's equilibrium radius $R_{\rm eq}= R(R_0)$ satisfies~$R_{\rm eq}\geq R_0$, and that the dominant energy condition is satisfied in the whole disk for all equilibria. The right panel of Figure~\ref{fig:Req} shows that, as might be anticipated, the dominant energy condition is saturated as the disk's edge velocity approaches the speed of light. Note that~$p_1$ is the principal pressure along~$\partial_{\lambda^1}$, and~$p_2$ along~$\partial_{\lambda^2}+\left[\Omega R^2/(1-\Omega^2R^2)\right] \partial_\tau$.

The rigid membrane is, of course, just one among infinitely many possible choices for the disk's elastic law. It has the advantage of being relatively simple, and is suggestive of astrophysical objects such as neutron star crusts, whose speeds of sound are believed to be relativistic \cite{AER22}. Less rigid materials, with smaller speeds of sound, would very likely lead to qualitatively similar results, but greater deformations.
%
%
%
%
\section{Dyson sphere}\label{section3}
In this section we consider a {\em Dyson sphere} \cite{Dyson60}, that is, a spherical membrane at equilibrium in Schwarzschild's spacetime (a situation which cannot occur for Nambu-Goto, or Dirac, membranes, see \cite{LL96}). We obtain the equilibrium configuration, where the gravitational attraction is balanced by the isotropic tangential pressure, and analyze its linear stability. Surprisingly, although the spherically symmetric mode is stable (yielding what is usually called a {\em breathing mode}), the axi-symmetric dipolar mode is already unstable. This may be taken as a cautionary tale against misrepresenting radial stability as true stability.
\subsection{Equilibrium conditions}
Let us start with the Schwarzschild solution of mass $M > 0$, given in Schwarzschild coordinates by
\begin{equation}\label{S_metric}
ds^2 = - \left(1 - \frac{2M}{r} \right) dt^2 + \left(1 - \frac{2M}{r} \right)^{-1} dr^2 + r^2 \left(d \theta^2 + \sin^2 \theta d\varphi^2\right) \, .
\end{equation}
Consider a spherical membrane of radius $R_0$, corresponding to the relaxed metric
\begin{equation}
k = \left(d\lambda^1\right)^2 + \sin^2\left( \frac{\lambda^1}{R_0}\right)\left(d\lambda^2\right)^2 ,
\end{equation}
with $(\lambda^1, \lambda^2) \in \left(0,\pi R_0\right) \times \left(0,2\pi R_0\right)$. As discussed in Section~\ref{section1}, the motion of this membrane in the Schwarzschild geometry is described by an embedding, which we take to be of the form
\begin{equation} \label{embed}
\begin{cases}
t(\tau,\lambda^1,\lambda^2)=\tau \\
r(\tau,\lambda^1,\lambda^2) = R \\
\theta(\tau,\lambda^1,\lambda^2) = \lambda^1/R_0 \\
\varphi(\tau,\lambda^1,\lambda^2) = \lambda^2/R_0
\end{cases},
\end{equation}
corresponding to the shell being placed at rest at $r=R$. The metric induced by the embedding on the worldtube is
\begin{equation}
(h_{AB}) = 
\left(
\begin{matrix}
- \left(1 - \frac{2M}{R} \right) & 0 & 0 \\
0 & \frac{R^2}{R_0^2} & 0 \\
0 & 0 & \frac{R^2}{R_0^2} \sin^2\left( \frac{\lambda^1}{R_0}\right)
\end{matrix}
\right) \, ,
\end{equation}
and so
\begin{equation} \label{n^2}
n^2 = \frac{h_{00}f^2}{h} =  \frac{R_0^4}{R^4}
\end{equation}
and
\begin{equation} \label{s^2}
s^2 = h_{11} + \frac{h_{22}}{f^2} - \frac1{h_{00}} \left( h_{01}^2 + \frac{h_{02}^2}{f^2} \right) = \frac{2R^2}{R_0^2} \, .
\end{equation}
Moreover, we have
\begin{equation}
(h^{AB}) = 
\left(
\begin{matrix}
- \left(1 - \frac{2M}{R} \right)^{-1} & 0 & 0 \\
0 & \frac{R_0^2}{R^2} & 0 \\
0 & 0 & \frac{R_0^2}{R^2} \sin^{-2}\left( \frac{\lambda^1}{R_0}\right)
\end{matrix}
\right)
\end{equation}
and
\begin{equation}
(k^{AB}) = 
\left(
\begin{matrix}
0 & 0 & 0 \\
0 & 1 & 0 \\
0 & 0 & \sin^{-2}\left( \frac{\lambda^1}{R_0}\right)
\end{matrix}
\right) \, ,
\end{equation}
so that
\begin{equation}
(T^{AB}) = 
\left(
\begin{matrix}
\left(1 - \frac{2M}{R} \right)^{-1} F & 0 & 0 \\
0 & \frac{R_0^2}{R^2}\left( 2 n^2 F^{(1,0)} - F \right) - 2F^{(0,1)} & 0 \\
0 & 0 & \left[ \frac{R_0^2}{R^2}\left( 2 n^2 F^{(1,0)} - F \right) - 2F^{(0,1)} \right] \sin^{-2}\left( \frac{\lambda^1}{R_0}\right)
\end{matrix}
\right)  \, .
\end{equation}
After a long but straightforward computation, the equations of motion \eqref{motion} boil down to
\begin{equation} \label{S_eq}
\frac{M}{R^2} F = 2 \left( 1-\frac{2M}{R} \right) \left[ \frac{1}{R} \left( 2n^2 F^{(1,0)} - F \right) + \frac{2R}{R_0^2} F^{(0,1)}\right] \, ,
\end{equation}
which, in view of~\eqref{rhoandp} and \eqref{s^2}, can be written in the suggestive form
\begin{equation} \label{S_eq2}
\frac{M \rho}{R} = 2 \left( 1-\frac{2M}{R} \right) p \, .
\end{equation}
In fact, this is simply the generalized sail equation \eqref{constraints}; the conservation equations \eqref{conservation2} hold automatically for this embedding. For comparison, the Newtonian equation for a spherical membrane with uniform density $\rho$ and isotropic tangential pressure $p$ in the field of a point mass $M$ placed at its center is
\begin{equation} \label{Newton_eq}
\frac{M \rho}{R} = 2 p \, .
\end{equation}
\subsection{Linear Stability}
To analyse the linear stability of the Dyson sphere we consider the embedding
\begin{equation} \label{embed2}
	\begin{cases}
		t(\tau,\lambda)=\tau \\
		r(\tau,\lambda) = R + \delta r(\tau, \lambda^1, \lambda^2)\\
		\theta(\tau, \lambda) = \lambda^1/R_0+\delta \theta(\tau, \lambda^1, \lambda^2) \\
		\varphi(\tau,\lambda) = \lambda^2/R _0+ \delta \varphi(\tau, \lambda^1, \lambda^2)
	\end{cases}
\end{equation}
around an equilibrium configuration (satisfying \eqref{S_eq}). Substituting \eqref{embed2} in \eqref{S_metric} we obtain, to first order,
\begin{equation*}
(h_{AB}) = 
\left(
\footnotesize\begin{matrix}
-1+\frac{2M}{R}-\frac{2M}{R^2} \delta r & \frac{R^2}{R_0} \delta \dot{\theta}  &  \frac{R^2}{R_0} \sin^2\left( \frac{\lambda^1}{R_0}\right) \delta \dot{\varphi} \\
\frac{R^2}{R_0} \delta \dot{\theta} & \frac{R^2}{R_0^2} + \frac{2R}{R_0^2} \delta r + \frac{2R^2}{R_0}\delta \theta_1 & \frac{R^2}{R_0}\delta \theta_2 +  \frac{R^2}{R_0} \sin^2\left( \frac{\lambda^1}{R_0}\right) \delta \varphi_1 \\
\frac{R^2}{R_0} \sin^2\left( \frac{\lambda^1}{R_0}\right) \delta \dot{\varphi} & \frac{R^2}{R_0}\delta \theta_2 +  \frac{R^2}{R_0} \sin^2\left( \frac{\lambda^1}{R_0}\right) \delta \varphi_1 & \frac{R(R+2\delta r)}{R_0^2} \sin^2\left( \frac{\lambda^1}{R_0}\right) +  \frac{R^2}{R_0^2} \sin\left( \frac{2\lambda^1}{R_0}\right)\delta\theta + \frac{2R^2}{R_0} \sin^2\left( \frac{\lambda^1}{R_0}\right) \delta \varphi_2 
\end{matrix}
\right),
\end{equation*}
where $\dot{\phi} \equiv \frac{\partial \phi}{\partial \tau}$, $\phi_1= \frac{\partial \phi}{\partial \lambda^1}$ and $\phi_2= \frac{\partial \phi}{\partial \lambda^2}$ for any function $\phi(\tau, \lambda^1, \lambda^2)$.
Therefore, substituting in \eqref{TAB} and \eqref{motion}, we obtain the linearized equations of motion:
\begin{align}
	&\delta \ddot{r}+\left[(c_{L}^2-c_{TT}^2)\left(\tfrac{12M^2}{R^2}-\tfrac{14M}{R}+4\right)+\tfrac{3M^2}{R^2}-\tfrac{3M}{R}\right]\tfrac{\delta r}{R^2} +\tfrac{M R_0^2}{2 R^3}\Big[\delta r_{11}+\csc^2(\tfrac{\lambda^1}{R_0})\,\delta r_{22}+\cot(\tfrac{\lambda^1}{R_0}) \tfrac{\delta r_1}{R_0}\Big]\nonumber \\
	&+\left[(c_L^2-c_{TT}^2)\big(1-\tfrac{2M}{R}\big)-\tfrac{M}{2R} \right] \big(2-\tfrac{3 M}{R}\big) \left[\tfrac{R_0}{R}(\delta \theta_1+\delta \varphi_2)+\cot (\tfrac{\lambda^1}{R_0})\tfrac{\delta \theta}{R}\right] =0 \, ,\\
	&\delta \ddot{\theta}+\big(1-\tfrac{2M}{R}\big)\left[(c_L^2-c_{TT}^2)+c_{TT}^2 \cos(\tfrac{2\lambda^1}{R})\right]\csc^2(\tfrac{\lambda^1}{R_0})\tfrac{\delta \theta}{R^2}- \tfrac{c_L^2 R_0^2}{R^2}\big(1-\tfrac{2M}{R}\big)\Big[\delta \theta_{11}+\cot (\tfrac{\lambda^1}{R_0}) \tfrac{\delta \theta_1}{R_0}+\tfrac{c_{TT}^2}{c_L^2}\csc^2(\tfrac{\lambda^1}{R_0})\delta \theta_{22}\Big]\nonumber \\
	&-\tfrac{2R_0}{R^3}\left[(c_L^2-c_{TT}^2)\big(1-\tfrac{2M}{R}\big)-\tfrac{M}{2R} \right]\delta r_1-\tfrac{R_0^2}{R^2}\big(1-\tfrac{2 M}{R}\big)(c_L^2-c_{TT}^2)\delta \varphi_{12}+\tfrac{2c_{TT}^2 R_0 }{R^2}\big(1-\tfrac{2M}{R}\big)\cot (\tfrac{\lambda^1}{R_0})\delta \varphi_2=0 \, , \\
	&\delta \ddot{\varphi}-\tfrac{R_0^2}{R^2}\big(1-\tfrac{2M}{R}\big)[c_L^2\csc^2(\tfrac{\lambda^1}{R_0}) \delta \varphi_{22}+c_{TT}^2 (\delta \varphi_{11}+3 \cot(\tfrac{\lambda^1}{R_0})\tfrac{\delta \varphi_1}{R_0})]-\tfrac{2R_0}{R^3}\left[(c_L^2-c_{TT}^2)\big(1-\tfrac{2M}{R}\big)-\tfrac{M}{2R} \right]\csc^2(\tfrac{\lambda^1}{R_0})\delta r_2\nonumber \\
	&-\tfrac{R_0}{R^2}(c_L^2+c_{TT}^2)\big(1-\tfrac{2 M}{R}\big)\cot (\tfrac{\lambda^1}{R_0})\csc ^2(\tfrac{\lambda^1}{R_0})\delta \theta_2-(c_L^2-c_{TT}^2)\tfrac{R_0^2}{R^2}\big(1-\tfrac{2 M}{R}\big)\csc^2(\tfrac{\lambda^1}{R_0})\delta \theta_{12}=0 \, .
\end{align}

Because $p>0$, we have $c_T^2<0$, and so we expect instabilities at small length scales (as noted in \cite{Carter89b} in the case of strings). To examine the behavior of the Dyson sphere at large length scales we examine the so-called {\em breathing mode}, corresponding to spherically symmetric deformations, that is, $\delta r=\delta r(\tau)$, $\delta \theta=0$ and $\delta \varphi=\delta \varphi(\tau)$. The linearized equations of motion yield in this case $\delta \ddot{\varphi}=0$ and
\begin{equation}
\delta \ddot{r}+\left[(c_{L}^2-c_{TT}^2)\left(\tfrac{12M^2}{R^2}-\tfrac{14M}{R}+4\right)+\tfrac{3M^2}{R^2}-\tfrac{3M}{R}\right]\tfrac{\delta r}{R^2} =0 \, .
\end{equation}
This yields the stability condition
\begin{equation}\label{stability}
c_{L}^2-c_{TT}^2 > \frac{\tfrac{3M}{R}-\tfrac{3M^2}{R^2}}{4-\tfrac{14M}{R}+\tfrac{12M^2}{R^2}}  \, ,
\end{equation}
which is easily satisfied by physically reasonable elastic membranes for $R \gg M$, as it approaches the condition that the bulk modulus should be positive, $c_{L}^2-c_{TT}^2 >0$. For smaller values of $R$ this condition becomes harder to satisfy, and in fact the right-hand side of \eqref{stability} is equal to $1$ for $R=3M$, so that no physical Dyson sphere can supports a breathing mode for $R \leq 3M$.

Going one step beyond the breathing mode, we consider axisymmetric perturbations, corresponding to $\delta r=\delta r(\tau,\lambda^1)$, $\delta \theta=\delta \theta(\tau, \lambda^1)$ and $\delta \varphi=0$. The linearized equations of motion yield in this case
\begin{align}
	&\delta \ddot{r}+\left[(c_{L}^2-c_{TT}^2)\left(\tfrac{12M^2}{R^2}-\tfrac{14M}{R}+4\right)+\tfrac{3M^2}{R^2}-\tfrac{3M}{R}\right]\tfrac{\delta r}{R^2} +\tfrac{M R_0^2}{2 R^3}\Big[\delta r_{11}+\cot(\tfrac{\lambda^1}{R_0}) \tfrac{\delta r_1}{R_0}\Big]\nonumber \\
	&+\left[(c_L^2-c_{TT}^2)\big(1-\tfrac{2M}{R}\big)-\tfrac{M}{2R} \right] \big(2-\tfrac{3 M}{R}\big) \left[\tfrac{R_0}{R}\delta \theta_1+\cot (\tfrac{\lambda^1}{R_0})\tfrac{\delta \theta}{R}\right] =0 \, ,\\
	&\delta \ddot{\theta}+\big(1-\tfrac{2M}{R}\big)\left[(c_L^2-c_{TT}^2)+c_{TT}^2 \cos(\tfrac{2\lambda^1}{R})\right]\csc^2(\tfrac{\lambda^1}{R_0})\tfrac{\delta \theta}{R^2}\nonumber \\
	&- \tfrac{c_L^2 R_0^2}{R^2}\big(1-\tfrac{2M}{R}\big)\Big[\delta \theta_{11}+\cot (\tfrac{\lambda^1}{R_0}) \tfrac{\delta \theta_1}{R_0}\Big]-\tfrac{2R_0}{R^3}\left[(c_L^2-c_{TT}^2)\big(1-\tfrac{2M}{R}\big)-\tfrac{M}{2R} \right]\delta r_1=0 \,  .
\end{align}
Setting
\begin{align}
& \delta r(\tau,\lambda^1) = \alpha(\tau) \cos(\tfrac{\lambda^1}{R_0}) \, , \\
& \delta \theta(\tau,\lambda^1) = \beta(\tau) \sin(\tfrac{\lambda^1}{R_0}) \, ,
\end{align}
we obtain
\begin{align}
	&\ddot{\alpha}+\left[(c_{L}^2-c_{TT}^2)\left(\tfrac{12M^2}{R^2}-\tfrac{14M}{R}+4\right)+\tfrac{3M^2}{R^2}-\tfrac{4M}{R}\right]\tfrac{\alpha}{R^2} \nonumber \\
	&\qquad \qquad \qquad +2\left[(c_L^2-c_{TT}^2)\big(1-\tfrac{2M}{R}\big)-\tfrac{M}{2R} \right] \big(2-\tfrac{3M}{R}\big)\tfrac{\beta}{R} =0 \, ,\\
	&\ddot{\beta}+2\big(1-\tfrac{2M}{R}\big)(c_L^2-c_{TT}^2)\tfrac{\beta}{R^2}+2\left[(c_L^2-c_{TT}^2)\big(1-\tfrac{2M}{R}\big)-\tfrac{M}{2R} \right]\tfrac{\alpha}{R^3}=0 \,  .
\end{align}
Now the stability requires that the matrix
\begin{equation}
\Omega =
\left(
\begin{matrix}
\left[(c_{L}^2-c_{TT}^2)\left(\tfrac{12M^2}{R^2}-\tfrac{14M}{R}+4\right)+\tfrac{3M^2}{R^2}-\tfrac{4M}{R}\right]\tfrac{1}{R^2} &
2\left[(c_L^2-c_{TT}^2)\big(1-\tfrac{2M}{R}\big)-\tfrac{M}{2R} \right] \big(2-\tfrac{3M}{R}\big)\tfrac{1}{R} \\
2\left[(c_L^2-c_{TT}^2)\big(1-\tfrac{2M}{R}\big)-\tfrac{M}{2R} \right]\tfrac{1}{R^3} & 2\big(1-\tfrac{2M}{R}\big)(c_L^2-c_{TT}^2)\tfrac{1}{R^2}
\end{matrix}
\right)
\end{equation}
has positive eigenvalues. However, we have
\begin{equation}
\det \Omega = \tfrac{M^2}{R^6} \left[-2 + \tfrac{3M}{R} - 6 (c_L^2-c_{TT}^2) \big(1-\tfrac{2M}{R}\big) \right] < 0
\end{equation}
for
\begin{equation}
c_L^2-c_{TT}^2 > - \frac{(2 - \tfrac{3M}{r})}{6 (1 - \tfrac{2M}{R})} \, ,
\end{equation}
which holds for any physically reasonable elastic membrane. Therefore, we have an instability of the Dyson sphere already at the level of the dipolar mode.
%
%
\section{Conclusions}
To conclude, we briefly summarize our results and discuss future directions of research. 

\subsection{Formalism}

In the first part of this work we (re-)derived the equations of motion for elastic membranes starting from a Lagrangian density, and recast these equations as conservation of energy-momentum along the worldtube plus the {\em generalized sail equations} (the vanishing of the contraction between the energy-momentum tensor and the extrinsic curvature of the worldtube). We also obtained the conserved quantities in spacetimes with Killing vector fields and computed the membrane's longitudinal and transverse speeds of sound. Finally, we determined the membrane's coefficients of linear elasticity: the Poisson ratio, the bulk modulus, the shear modulus and the Young modulus.

\subsection{Rigidly rotating disk}

In the second part of this work we gave an example of a simple $1$-parameter family of elastic laws with longitudinal speed of sound equal to the speed of light, which we dubbed the {\em rigid membrane}.  We determined its transverse speeds of sound, and computed its coefficients of linear elasticity with respect to the relaxed configuration. We then used this elastic law to produce, for the first time, explicit examples of fully relativistic rigidly rotating disks, depicted in Figure~\ref{fig:Req}.

It would be interesting to conduct a linear stability analysis of these rigidly rotating disks. Additionally, obtaining analogous solutions for rotating annuli in the equatorial plane of the Kerr spacetime (along with a study of their linear stability) could provide simplified models for thin accretion disks.

\subsection{Dyson sphere}

In the third part of this work we considered a {\em Dyson sphere}, that is, a spherical membrane at equilibrium in Schwarzschild's spacetime. We obtained the equilibrium configuration, where the gravitational attraction is balanced by the isotropic tangential pressure, and analyzed its linear stability. We found that while the spherically symmetric mode is stable, the axi-symmetric dipolar mode is already unstable. This warns against misrepresenting radial stability as true stability.

It would be interesting to consider a rotating Dyson sphere, which could be under tension, thus avoiding the short wavelength instability associated with an imaginary tranverse speed of sound. Would such a Dyson sphere be linearly stable?
%
%
\section*{Acknowledgements}
We thank Martin Kolo\v{s} for his suggestion of studying Dyson spheres. P.M.\ and J.N.\ were partially supported by Funda\c{c}\~ao para a Ci\^encia e a Tecnologia (Portugal) through CAMGSD, IST-ID (projects UIDB/04459/2020 and UIDP/04459/2020), and also by the H2020-MSCA-2022-SE project EinsteinWaves, GA no.~101131233. R.V. was supported by grant no.~FJC2021-046551-I, funded by MCIN/AEI/10.13039/501100011033, and by the European Union NextGenerationEU/PRTR. R.V.\ also acknowledges the support from the Departament de Recerca i Universitats from Generalitat de Catalunya to the Grup de Recerca `Grup de F\'isica Te\`orica UAB/IFAE' (Codi: 2021 SGR 00649).
%
%

\end{document}